\documentclass[letterpaper,twocolumn,10pt]{article}
\usepackage{usenix,epsfig,endnotes,amsmath,multirow,mathtools}
\usepackage{changepage}
\usepackage{xcolor}
\usepackage{todonotes}
\usepackage[ruled]{algorithm2e}
\usepackage{hyperref}
\usepackage{caption}
\usepackage{subcaption}

\newcommand{\rk}[1]{{\color{red} #1}}

\begin{document}

\date{}

\title{\Large \bf Image Obfuscation for Privacy-Preserving Machine Learning}

\author{
{\rm Mathilde Raynal}\\
Cyber-Defence Campus \\ armasuisse S+T \\ Switzerland
\and
{\rm Radhakrishna Achanta}\\
Swiss Data Science Center \\ EPFL and ETH Zurich \\ Switzerland
\and
{\rm Mathias Humbert}\\
Cyber-Defence Campus \\ armasuisse S+T \\ Switzerland
}

\maketitle

\thispagestyle{empty}

\section*{Abstract}
Privacy becomes a crucial issue when outsourcing the training of machine learning (ML) models to cloud-based platforms offering machine-learning services. While solutions based on cryptographic primitives have been developed, they incur a significant loss in accuracy or training efficiency, and require modifications to the backend architecture. A key challenge we tackle in this paper is the design of image obfuscation schemes that provide enough privacy without significantly degrading the accuracy of the ML model and the efficiency of the training process. In this endeavor, we address another challenge that has persisted so far: quantifying the degree of privacy provided by visual obfuscation mechanisms. We compare the ability of state-of-the-art full-reference quality metrics to concur with human subjects in terms of the degree of obfuscation introduced by a range of techniques. By relying on user surveys and two image datasets, we show that two existing image quality metrics are also well suited to measure the level of privacy in accordance with human subjects as well as AI-based recognition, and can therefore be used for quantifying privacy resulting from obfuscation. With the ability to quantify privacy, we show that we can provide adequate privacy protection to the training image set at the cost of only a few percentage points loss in accuracy. 

\section{Introduction}\label{sec:intro}

The recent breakthroughs in machine learning, notably through deep learning, have gone hand-in-hand with an extraordinary increase in the amount of data and computation to train complex models that have unprecedented number of parameters. This, in turn, has given rise to several commercial platforms that offer Machine Learning as a Service (MLaaS), which easily scale up to the wide ranging needs of memory and computational power of different applications. A client can submit a labeled dataset to such a service and obtain in return a trained model for a given machine-learning application.

While both efficient and cost effective, such services have given rise to the need for protecting the privacy of the data that has to leave the client's premises. Providing raw training data is not just a matter of trusting the service, it may even be legally restricted, depending on how sensitive the data is considered. A common approach for enhancing privacy in this context is to encrypt the data and rely on cryptographic schemes such as homomorphic encryption~\cite{aono2016scalable, mlhe16, bonte2018privacy, fhe18priv, graepel2012ml, kim2018secure, nandakumar2019towards}. While strong on privacy, current methods for training on encrypted data have serious limitations on the type of ML model they can handle or on their resulting accuracy. Moreover, this approach introduces a non-negligible computational overhead on training, while at the same time necessitating drastic modifications to the backend. 

Another popular approach is to apply noise on the input data directly, output data, or on the gradients in a differentially private manner~\cite{ACG+16,CDN+15,fredrikson2014privacy,huang2019dp,jayaraman2019evaluating,lee2019synthesizing, SS15}. However, this approach often brings strong privacy guarantees at the detriment of accuracy, especially when it is applied to raw input data~\cite{F18,F19,lee2019synthesizing}. As we can see, there an inherent privacy-utility trade-off in privacy-preserving ML model training. In this paper, we challenge this inherent trade-off by presenting new methods to enhance the privacy of the training data while maximizing the resulting model accuracy and without adding  computational complexity to the training process. One key challenge then is to be able to choose the appropriate degree of privacy since this is not trivially quantifiable. This becomes even harder when the data is of audio-visual nature because the quantification has to agree with human perception. We propose a principled way to quantify the privacy introduced, specifically in the case of images, through the use of visual obfuscation and image quality metrics. Based on our image privacy metrics, we then show how we can trade-off the accuracy of a typical deep learning model to the degree of privacy we seek.

\textbf{Contributions.} We introduce new obfuscation mechanisms for images by combining various well-established methods with \emph{mixup}, a  technique that consists of linearly mixing image samples and their labels as a means of regularizing deep learning models~\cite{ZCD+18}. We then seek state-of-the-art full-reference image quality metrics (IQM) to quantify the degree of \emph{degradation} of the image. In order to assess the ease of recognition by humans and the relevance of the metrics, we design and perform a user survey by using two different image datasets that we obfuscate with our new obfuscation schemes. Through this objective comparison, we discover the metrics that agree most with human perception, namely SSIM~\cite{WBS+04} and HaarPSI~\cite{RBK+18}. Moreover, thanks to these metrics, we are able to define numerical thresholds on the degree of image degradation that provides adequate privacy. Note that these thresholds do not depend on the image obfuscation technique. Our thresholds simply indicate by how much a given images is to be degraded according to the corresponding image quality metric so that it can no longer be visually understood, thereby quantifying image privacy.

We additionally confirm the efficacy of our method for quantifying image privacy by testing the accuracy of the Google Vision AI\footnote{\url{https://cloud.google.com/vision}} in recognizing content in the image. The results confirm that we can successfully choose the right degree of degradation to conceal content from both human visual analysis and artificial recognition systems. By extension, the methodology should be applicable for other applications as well as other signals.
To the best of our knowledge, there is no previous work that quantifies image obfuscation level as we propose. We further study other attack vectors and notably show that the proposed obfuscation mechanisms (with appropriate privacy parameters) dramatically decrease the success of the deep learning-based reconstruction attack proposed by McPherson et al.~\cite{MSS16}.

Finally, we rely on the privacy metrics defined above to evaluate the trade-off between privacy and utility in the different obfuscation mechanisms that we propose and compare them. We show that \emph{noise and mixing} and \emph{shuffling and mixing} provide the best trade-off between privacy and the accuracy of the deep learning model trained on the obfuscated images. Specifically, for privacy values above the optimal threshold derived from human perception, \emph{noise and mixing} provides an accuracy that is only a few percentage points lower than the accuracy on non-obfuscated images. 

We summarize our contributions as follows:

\begin{itemize}
    \item We provide a methodical way to quantify privacy attained through visual obfuscation in images.
    \item We assess the efficacy of several existing full-reference image quality metrics in quantifying privacy by re-purposing them for this task.
    \item We propose and assess several obfuscation techniques to preserve privacy without severely deteriorating the accuracy of the trained models.
    \item Finally, we perform an in-depth analysis of the relation between the parameters of the obfuscation techniques and their privacy-utility trade-off.
\end{itemize}

The rest of the paper is organized as follows.
The adversarial model is described in Section~\ref{sec:threats}.
Section~\ref{sec:obf} introduces our obfuscation techniques.
Section~\ref{sec:privacy} explains how we evaluate different image quality metrics for privacy.
Section~\ref{sec:results} brings it all together and delivers the utility-privacy results.
Related work is presented in Section~\ref{sec:related}.
Section~\ref{sec:conclusion} concludes the paper along with suggestions for future work.

\section{Threat Model}\label{sec:threats}

We assume the adversary has access to the images of a training set used to construct a machine-learning model. This adversary is typically the cloud service provider such as the MLaaS provider to which the client will outsource the training of its machine-learning model. Similarly, if the data is released publicly, e.g., for research purposes, the adversary will be anyone having access to the released images. The adversary can either inspect the images visually or use some automated computer vision-based tool.
The adversary has access to the images shared by the client, i.e., the obfuscated images. We do not assume the adversary to have access to any auxiliary database containing the original images (before the obfuscation) since all images used in the obfuscation phase are considered to be private. 

In addition to the obfuscated images, the adversary has access to the label of one of the original images included in the obfuscated image, as discussed in Section~\ref{sec:obf}. The labels are not considered sensitive by themselves. The premise is that, e.g., in an MRI image classification problem,  revealing whether a training image contains a tumor or not is not as sensitive as revealing the personally identifiable information such as their gender or age, or any other disease that is not part of the classification task. Inferring other information than the label is equivalent to property inference as defined in~\cite{ateniese2015hacking,ganju2018property, MSC+19} but at the individual data level instead of aggregate.

Other works have studied privacy attacks with black-box access to the machine-learning models~\cite{fredrikson2015model,fredrikson2014privacy,hayes2019logan,salem2019ml,shokri2017membership,tramer2016stealing}. In our case, the adversary is assumed to be much stronger, by having directly access to the raw -- yet obfuscated -- training data. 
From this point of view, our work is complementary to previous papers and could help mitigate some of the known black-box attacks. For instance, our proposed obfuscation techniques would add a layer of protection against model inversion attacks where the adversary aims to reconstruct the original training data~\cite{fredrikson2015model}.
\section{Obfuscation Techniques}\label{sec:obf}
In this section, we introduce the different obfuscation techniques we study to enhance image privacy. All of the methods are variants (see also Figure~\ref{fig:obfuscated}) of the \emph{mixup} technique introduced by Zhang et al.~\cite{ZCD+18}. We rely on mixup due to its intriguing capability of increasing the accuracy of deep learning models used for image classification despite deliberately introducing visual artifacts in images. Mixup involves constructing convex combinations of pairs of examples drawn at random from the training data as follow:
$$\tilde{x} = \lambda x_i + (1 - \lambda)x_j$$
$$\tilde{y} = \lambda y_i + (1 - \lambda)y_j$$ 
where $x_i$, $x_j$ are raw input vectors, $y_i$, $y_j$ are one-hot label encodings and $\lambda \in [0, 1]$ is drawn from a beta distribution.
As new combinations are created at every epoch, mixup acts as a data augmentation technique and helps regularizing the learning of the model.
We refer to the original paper~\cite{ZCD+18} for the exhaustive list of the benefits of mixup.

\subsection{Mixing}
The initial mixup algorithm consists of a combination of two images with weights $\lambda$ and (1-$\lambda$), where $\lambda$ is randomly generated from a beta distribution. The hyperparameters of the beta distribution have to be fine-tuned to fit the underlying task.
In the case of image classification on the ImageNet-2012 dataset, sampling $\lambda$ from $Beta(\alpha, \alpha)$ with $\alpha \in \{0.2, 0.4\}$ is shown to lead to improved accuracy.

However, this approach cannot guarantee maximal visual obfuscation of every mixture as $\lambda$ has a high probability to be close to 0 or 1.
In this work, we propose to define the $\lambda$ coefficient such that it maximizes the distortion of the images. Intuitively, $\lambda=0.5$ should achieve the maximal distortion. The weight $\lambda_i$ of the image $x_i$ will then be expressed as a function of the original images to combine: $\lambda_i =  M(x_i,  x_j)$.
As weights are constrained to sum to one, we impose that $ \lambda_j = \lambda(x_j, x_i) = 1 - \lambda_i$.

Inspired by the work of Inoue~\cite{Ino18}, we assign to the mixed image the label of the image with the larger contribution in the mixture instead of using a weighted combination of the two original labels. In case of $\lambda_i=0.5$, we randomly pick one of the two labels.
As we explain in Section~\ref{other_attacks}, not mixing the labels provide higher privacy guarantees while only slightly diminishing the data augmentation benefits of mixup on the accuracy. In our approach, the input and labels, respectively, can be synthesized as:
$$\tilde{x} = \lambda_i x_i + \lambda_j x_j$$
$$\tilde{y} = y_i \text{ if } \lambda_i \ge \lambda_j \text{ else } y_j$$


where $x_i$, $x_j$ are raw input vectors, $y_i$, $y_j$ are one-hot label encodings and $\lambda_i, \lambda_j \in [0, 1]$ are chosen such that $\lambda_i+\lambda_j=1$. Please note that mixing can easily be extended to more than two images by defining as many weights $\lambda_i$'s as the total number images. 

\subsection{Mixing and Pixel Grafting}
Grafting consists in replacing a proportion $p$ of the pixels of one image by the pixels of another. These pixels are randomly selected. We can interpret grafting as mixing at the level of individual pixels with $\lambda \in \{0,1\}$.

When using grafting only, the distinction between pixels of the two images used can often be distinct and easily detectable.
So it is easy for an attacker to separate the pixels into two sparse images, and then recover missing pixels by using basic interpolation techniques~\cite{S14,ZZG+17}. To mitigate this weakness and ensure that original images cannot be recovered, we combine the grafting method with image mixing.
Instead of grafting new pixels into an image, we first mix the pixels to be grafted with the other image and then perform the grafting step.
$$\tilde{x} = B \circ x_i +   \overline{B} \circ (\lambda x_i + (1-\lambda) x_j)$$
$$\tilde{y} = y_i \text{ if } p + (1-p)\lambda \ge \frac{1}{2} \text{ else } y_j$$ 
where $x_i$, $x_j$ are raw input vectors, $y_i$, $y_j$ are one-hot label encodings, $\lambda \in [0, 1]$  and $B$ a binary matrix with same dimension as the inputs, with a ratio of $p$ entries set to 1 and the rest to 0.
$\overline{B}$ denotes the complement of $B$, meaning that $B + \overline{B} = \mathbf{1}$, the matrix of all ones.
The Hadamard product $\circ$ is the element-wise product of matrices.

\subsection{Pixel Shuffling and Mixing}

Pixel shuffling is an effective way to remove structural information from an image.
When applied globally, shuffling is too destructive to allow for high quality training.
In order to limit the degree of distortion, we shuffle pixels locally in restricted neighborhoods of size $b\times b$.
The larger $b$ is, the closer we get to global shuffling.
As expected, the privacy introduced by local pixel shuffling causes a non-negligible drop in accuracy as the extent of shuffling increases.
We shuffle pixels in small neighbourhoods before mixing the images in order to introduce additional obfuscation to image mixture without significantly impacting the accuracy of the trained models.
With this approach, a new sample is:

$$\tilde{x} = \lambda S_i(x_i,b) + (1-\lambda) S_j(x_j,b)$$
$$\tilde{y} = y_i \text{ if } \lambda_i \ge \lambda_j \text{ else } y_j$$ 
where $x_i$, $x_j$ are raw input vectors, $y_i$, $y_j$ are one-hot label encodings, $\lambda_i, \lambda_j \in [0, 1]$ are chosen such that $\lambda_i+\lambda_j=1$ and $S(.,b)$ is a function that shuffles pixels in local neighborhoods of $b\times b$ pixel using freshly generated random permutations.

\subsection{Adding Noise and Mixing}
Modern image classifiers are not easily perturbed by random noise despite the fact that it provides good visual obfuscation for the human eye. Nonetheless, adding too much noise can have a detrimental effect on utility.
In order to preserve both privacy and utility, we add some noise before the mixing step to increase its visual obfuscation. The noise is drawn from a Gaussian distribution $\mathcal{N}(0,\sigma^2)$ independently for each pixel and added to each of the two original images before mixing them. The new sample is defined as follows:

$$\tilde{x} = \lambda (x_i + z_i) + (1-\lambda) (x_j + z_j)$$
$$\tilde{y} = y_i \text{ if } \lambda_i \ge \lambda_j \text{ else } y_j$$ 
where $x_i$, $x_j$ are raw input vectors, $y_i$, $y_j$ are one-hot label encodings, $\lambda_i, \lambda_j \in [0, 1]$ are chosen such that $\lambda_i+\lambda_j=1$ and $z_i$ and $z_j$ are noise matrices independently sampled from the normal distribution $\mathcal{N}(0,\sigma^2)$ of mean 0 and standard variation $\sigma$.

\newcommand{\colwidth}{0.49\linewidth}
\newcommand{\gap}{@{\hspace{.5mm}}}

\begin{figure}[t!]
    \centering
        \begin{tabular}{c\gap c\gap}
            \includegraphics[width=\colwidth]{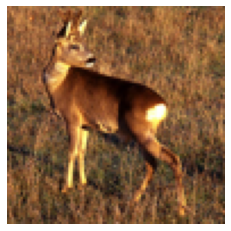} &  \includegraphics[width=\colwidth]{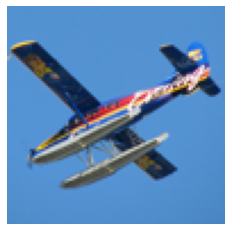}\\
            \multicolumn{2}{c}{{\small Original images for mixing}}\\
            \includegraphics[width=\colwidth]{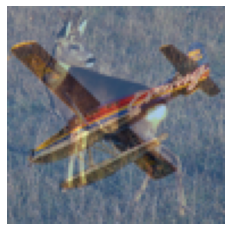} &  \includegraphics[width=\colwidth]{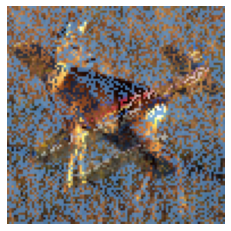}\\
            {\small Simple mixing} & {\small Mixing and grafting}\\
            \includegraphics[width=\colwidth]{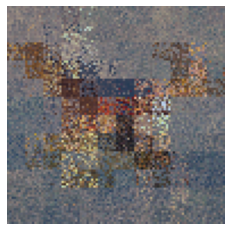} &  \includegraphics[width=\colwidth]{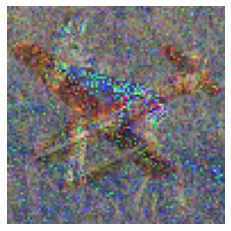}\\
             {\small Shuffling and mixing} & {\small Adding noise and mixing}\\
            \includegraphics[width=\colwidth]{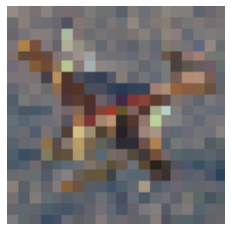} &  \includegraphics[width=\colwidth]{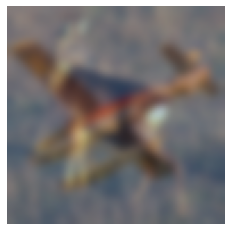}\\
            {\small Pixelizing and mixing} & {\small Blurring and mixing}\\
        \end{tabular}
    \caption{{Examples of obfuscated images created from an original image pair using the techniques mentioned in Section~\ref{sec:obf}.}}
    \label{fig:obfuscated}
\end{figure}

\subsection{Pixelizing and Mixing}
Pixelization can be used to obfuscate an image by lowering its resolution (although the actual size of the image stays the same).
The image is divided into a square grid, and each square is replaced by the average pixel value of the square.
The larger the square, the more pixels will be averaged together and the more detail will be lost. We pixelize the images before applying mixing on top to reach desirable privacy.
Thus, the final sample is:

$$\tilde{x} = \lambda R(x_i, b) + (1-\lambda) R(x_j, b)$$
$$\tilde{y} = y_i \text{ if } \lambda_i \ge \lambda_j \text{ else } y_j$$ 
where $x_i$, $x_j$ are raw input vectors, $y_i$, $y_j$ are one-hot label encodings, $\lambda_i, \lambda_j \in [0, 1]$ are chosen such that $\lambda_i+\lambda_j=1$ and $R(\cdot, b)$ a function that takes in input an image and pixelizes it with squares of size $b\times b$ pixels.

\subsection{Blurring and Mixing}
Blurring is widely used in graphics software, typically to reduce image noise or detail.
A Gaussian blur effect is typically performed by convolving an image with a Gaussian filter. However, as shown by previous works~\cite{LBH01,MSS16, NGB06}, perturbing the images with blurring alone is not an effective method to protect privacy.
As a consequence, we explore the technique consisting of combining both blurring and mixing by applying a blur operation before mixing the images.
The final sample is:
$$\tilde{x} = \lambda G(x_i, \sigma) + (1-\lambda) G(x_j, \sigma)$$
$$\tilde{y} = y_i \text{ if } \lambda_i \ge \lambda_j \text{ else } y_j$$ 
where $x_i$, $x_j$ are raw input vectors, $y_i$, $y_j$ are one-hot label encodings, $\lambda_i, \lambda_j \in [0, 1]$ are chosen such that $\lambda_i+\lambda_j=1$ and $G(\cdot, k)$ is a function that takes as input an image and applies a Gaussian blur with standard deviation $\sigma$. Note that blurring and mixing are commutative, thus blurring after mixing would result in the same obfuscated image.

\section{Quantifying Image Privacy}\label{sec:privacy}

In this section, we explain how we can quantify the degree of privacy introduced in an image via obfuscation. To this end, we compare state-of-the-art full-reference image quality metrics (IQM) for their agreement with the human visual system through subjective human tests to show how to derive scalar values that represent the degree of privacy. The framework we present is general enough to be extended to any new IQM for the purpose of quantifying privacy in images. To the best of our knowledge, it is the first such approach to quantify image privacy.

In the following, we first describe the various image similarity or distance metrics we rely on. Then, we evaluate all the metrics against human visual perception and computer recognition, and derive obfuscation thresholds for the metrics best capturing privacy. Finally, we discuss the limitations of these metrics and the robustness of our obfuscation methods in the face of other attack scenarios.

\subsection{Image Quality Metrics}
Full-reference image quality metrics (IQM) measure the difference between a reference image and its distorted version. While some of these metrics are claimed to be perceptually relevant, i.e.,  corresponding to the degree of distortion humans can see, it is a difficult claim to make. To understand this, one can consider the mean square error (MSE) metric, which is simply the Euclidean distance between the original and distorted images. If the distortion involved is a mere shift of the image by a single pixel, the change is imperceptible to the human eye. The MSE score, on the other hand, can be significantly high. So, it is well known that MSE is not perceptually relevant. This has led to the introduction of a plethora of other full-reference image quality metrics \cite{ HRU+16, KS08,liu2011image,RBK+18, metric06,WBS+04,zhang2014vsi} 
that try to bridge the gap between machine evaluation of image quality and its human perceptual relevance.

Here are the five image quality metrics that we evaluate in this paper:

\begin{itemize}
    \item MSE~\cite{GW92}: The mean squared error (MSE) measures the average squared difference over all pixels. In our case, the difference is computed between the obfuscated pixels and the original pixels. MSE is always non-negative and not upper-bounded.
    As we noted before, MSE might not be the most appropriate metric to quantify privacy as a simple one pixel shift might result in a high MSE score even though no privacy is introduced. We nonetheless include it to serve as the baseline metric.
    
    \item FID~\cite{HRU+16}: The Frechet Inception Distance (FID) is a metric initially used to evaluate the quality of images generated by generative adversarial networks.
    It captures the similarity between two images based on the deep features of the raw images calculated using the penultimate layer of an inception v3 model~\cite{szegedy2016rethinking}.
    Lower scores indicate the two images are more similar, or have more similar statistics, with a perfect score 0 indicating that the two images are identical.
    Its value is not upper-bounded.
        
    \item SSIM~\cite{WBS+04}: The structural similarity index measure (SSIM) is a method for predicting the perceived quality of all kinds of digital images based on an initial uncompressed or distortion-free image as reference.
    SSIM is a measure of spatial correlation consisting of three components: mean, variance, and cross-correlation. 
    It considers image degradation as perceived change in structural information, while also incorporating important perceptual phenomena, including both luminance masking and contrast masking terms.
    This measure accounts for the fact that spatially close pixels have strong interdependencies.
    SSIM score ranges from 0 to 1.
    We quantify privacy by reporting the dissimilarity dSSIM = 1-SSIM instead of the similarity.
    
    \item pHash~\cite{KS08}: A perceptual hash is a fingerprint of a multimedia file derived from various features of its content.
    Each hash bit is based on whether the pixel value is greater than the average frequency computed with a discrete cosine transform on the image.
    Perceptual hashes must be robust enough to take into account transformations or ``attacks'' on a given input and yet be flexible enough to distinguish between dissimilar files.
    Such attacks can include rotation, skew, contrast adjustment and different compression or formats. The distance measure used in this case is the normalized Hamming distance, which ranges from 0 to 1, between two binary fingerprints (hashes).

    \item HaarPSI~\cite{RBK+18}: The Haar wavelet-based perceptual similarity index (HaarPSI) is a similarity measure for images that aims to correctly assess the perceptual similarity between two images with respect to a human viewer.
    HaarPSI uses the coefficients obtained from a Haar wavelet decomposition to assess local similarities between two images, as well as the relative importance of image areas. The HarrPSI score ranges from 0 to 1.
    Similary to SSIM, HaarPSI is transformed into a dissimilarity (privacy) metric by computing dHaarPSI =  1-HaarPSI. We refer to it as dHaar for conciseness.
\end{itemize}

\subsection{User Perception}\label{sec:user}
In order to determine which metric is best suited for quantifying image obfuscation, and hence privacy, we need to compare the metrics with the actual perception of real human users. To do so, we conduct a user survey. We first present the image datasets and then the details of the survey.

\subsubsection{Building Obfuscated Image Samples}
We use two different image datasets for our survey: (i) the Stanford Dogs dataset~\cite{dogs} and (ii) the STL-10 dataset~\cite{stl10}.

The Stanford Dogs contains in total 120 breeds and 150 to 200 images per breed. In order to have a limited set of possible answers and recognizable breeds, we keep only 10 breeds. These breeds are chosen to be commonly known and, at the same time, have good breed recognition with automated systems~\cite{dogs}. The breeds we retain are the following: basset, german shepherd, golden retriever, husky, pembroke, poodle, pug, rottweiler, saint bernard, whippet.
As we need images of similar shapes for the mixing method, we equalize the size of all images to 256 $\times$ 256 pixels.

The STL-10 dataset is inspired by CIFAR-10 but has the notable advantage of containing higher resolution images ($96 \times 96$ pixels) as compared to CIFAR-10 ($32 \times 32$). This increases the recognition ability by human beings and thus does not artificially increase the original privacy that we would get with CIFAR-10 images because of low resolution. There are 10 classes in total: airplane, bird, car, cat, deer, dog, horse, monkey, ship, and truck. There are 500 images per class.

\subsubsection{User Survey}
We construct our obfuscated image samples by relying on the six different obfuscation techniques defined in Section~\ref{sec:obf}. Each technique appears in equal proportion throughout the survey, with the exception of mixing only, which is used either with two images or three images.
Each obfuscation technique is then used seven times, leading to 49 questions, i.e., images to recognize, in total.
The obfuscation level varies per question. The parameters are picked uniformly at random from a subset of values preselected per technique to provide interesting obfuscation, i.e., neither too easy nor too hard to guess.
The parameters and their possible values are listed in Table~\ref{table:params}.

\begin{table}[t]
\centering
\begin{tabular}{ | c | p{28mm}| } \hline
 Parameter & Values \\ \hline
 $\lambda_i$ & {0.5, 0.6, 0.7} \\ 
 $<\lambda>$ & (0.7, 0.2, 0.1), \newline (0.5, 0.33, 0.16), \newline (0.33, 0.33, 0.33) \\
 $p$ & {0.5, 0.6, 0.7, 0.8}\\
 $k$ & {17, 35, 45}\\
 $s$ & {16, 20, 32}\\
 $b$ & {4, 8, 16}\\
 $\sigma$ &  {10, 20, 40}\\ 
  \hline

\end{tabular}
 \caption{Possible parameter values for the obfuscation techniques used to generate the obfuscated images in the user survey. They are chosen such that they provide interesting obfuscated images, neither too easy nor too hard to guess.}
\label{table:params}
\end{table}

Each question contains one obfuscated image that is made from $n \in \{2, 3\}$ original images, depending on the obfuscation technique used.
To select the $n$ images to be used, $n$ classes are first selected, and then a random sample from that class is chosen.
By selecting the classes first, we ensure that we will have different labels in each obfuscated image.
All possible labels are presented to the user below each obfuscated image, along with an ``I cannot tell'' option.
We do not put a constraint on how many labels the user can choose, i.e., if the user recognizes the $n$ original images, she can answer the $n$ labels. But we emphasize the fact that, in case of doubt or hesitation between two labels, the ``I cannot tell'' option should be preferred.
Users are allowed to zoom in and out but not to perform any kind of modification on the image (contrast modification...) and take as much time as needed in order to simulate strong but realistic adversarial capabilities. For the Stanford Dogs-based survey, we include in the questionnaire one photo per breed as a reference to help the respondents recall the names of the breeds.
Sample pages of the questionnaires can be found in Appendix \ref{appendix:survey}.

\begin{figure*}[t]
    \centering
    \begin{tabular}{c c}
    \includegraphics[width=0.5\textwidth]{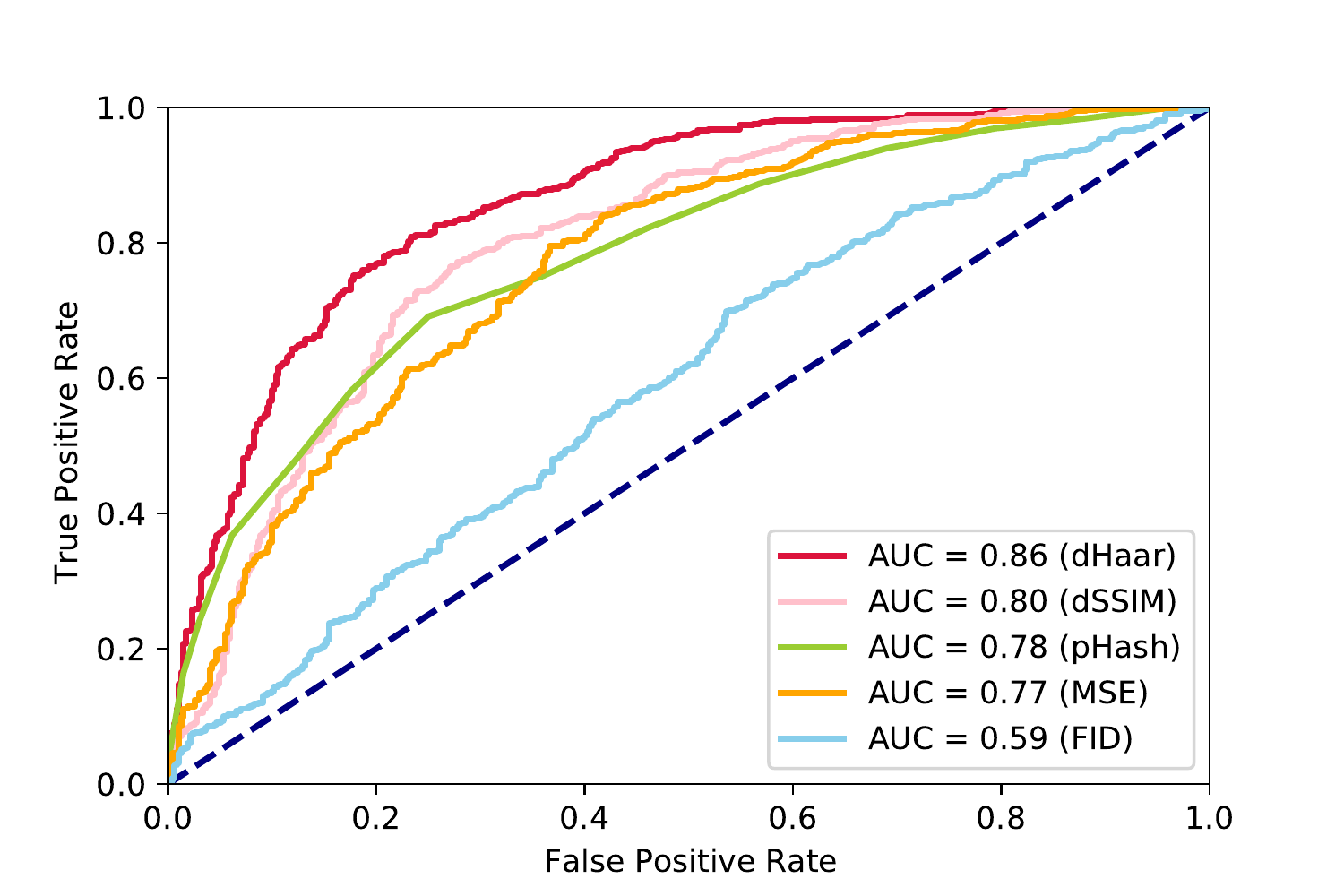} & \includegraphics[width=0.5\textwidth]{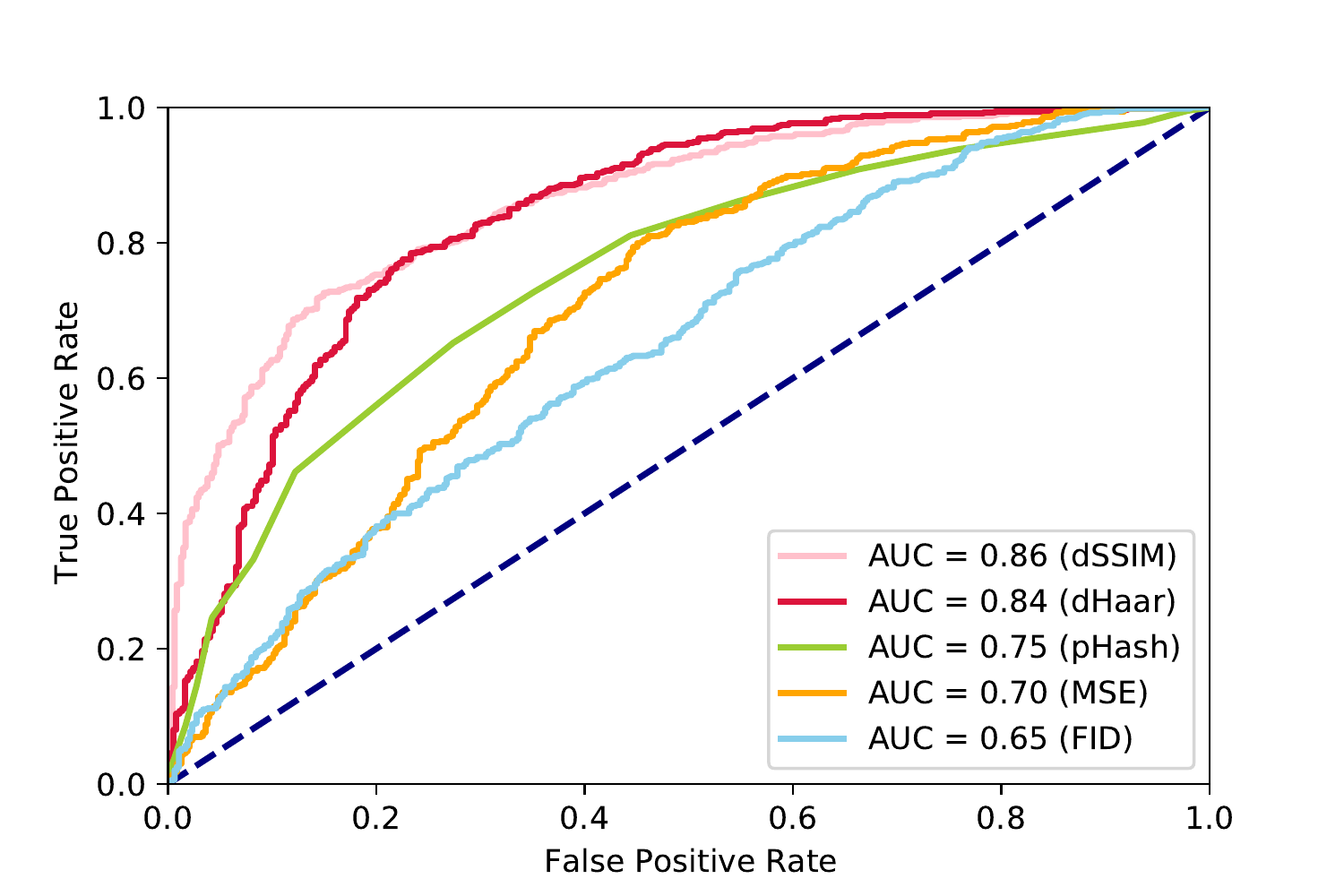}\\
    \end{tabular}
    \caption{ROC curves for different metrics against humans evaluations on the Stanford Dogs dataset (left) and STL-10 dataset (right). A higher curve indicates better agreement with human evaluations of the obfuscated images. As we can see, the dHaar and dSSIM metrics show better agreement with human subjects as compared to other metrics.}
    \label{fig:ROC}
\end{figure*}

\subsubsection{Survey Results}

We collected the results of 24 surveys in total, 12 per dataset.
This represents 1176 questions in total and 2544 pairwise comparisons (between the obfuscated images and the corresponding $n$ original images) that eventually allows to derive meaningful conclusions about the relation between the metrics and the human perception.
The 24 participants answering the surveys were mostly students (ranging from 23 to 34 years in age) with normal visual capabilities.

The goal of our experiment is (i) to identify which metric(s) fit(s) best human visual perception, and (ii) for each metric, what threshold leads to the best agreement between human perception and the obfuscation level given by the metric. In order to do so, we resort to receiver operating characteristic (ROC) curves~\cite{Faw06}. They are commonly used to evaluate the performance of binary classification systems by looking at the trade-off between the false-positive rate (FPR) and true-positive rate (TPR) for different threshold settings. We define as true positive the event where the metric is above the examined threshold and the human cannot recognize the image label. A false positive occurs when the metric is above this threshold yet the human can recognize the original image label.
Plotting the different trade-offs can be used to evaluate our different metrics.
The closer the curve is to the top-left corner -- (0,1) point -- of the graph, the better the ROC curve is deemed to be. In addition to the ROC curve, we rely on the area under the ROC curve (AUC), which provides a single measure to estimate the quality of the metrics.
The AUC ranges from 0 to 1, where a value of 0.5 is equivalent to random guessing, 0.7 to 0.8 is considered acceptable, 0.8 to 0.9 is considered excellent, and more than 0.9 is considered outstanding~\cite{HL00}.

\begin{figure*}[t]
    \centering
    \begin{tabular}{c c}
    \includegraphics[width=0.5\textwidth]{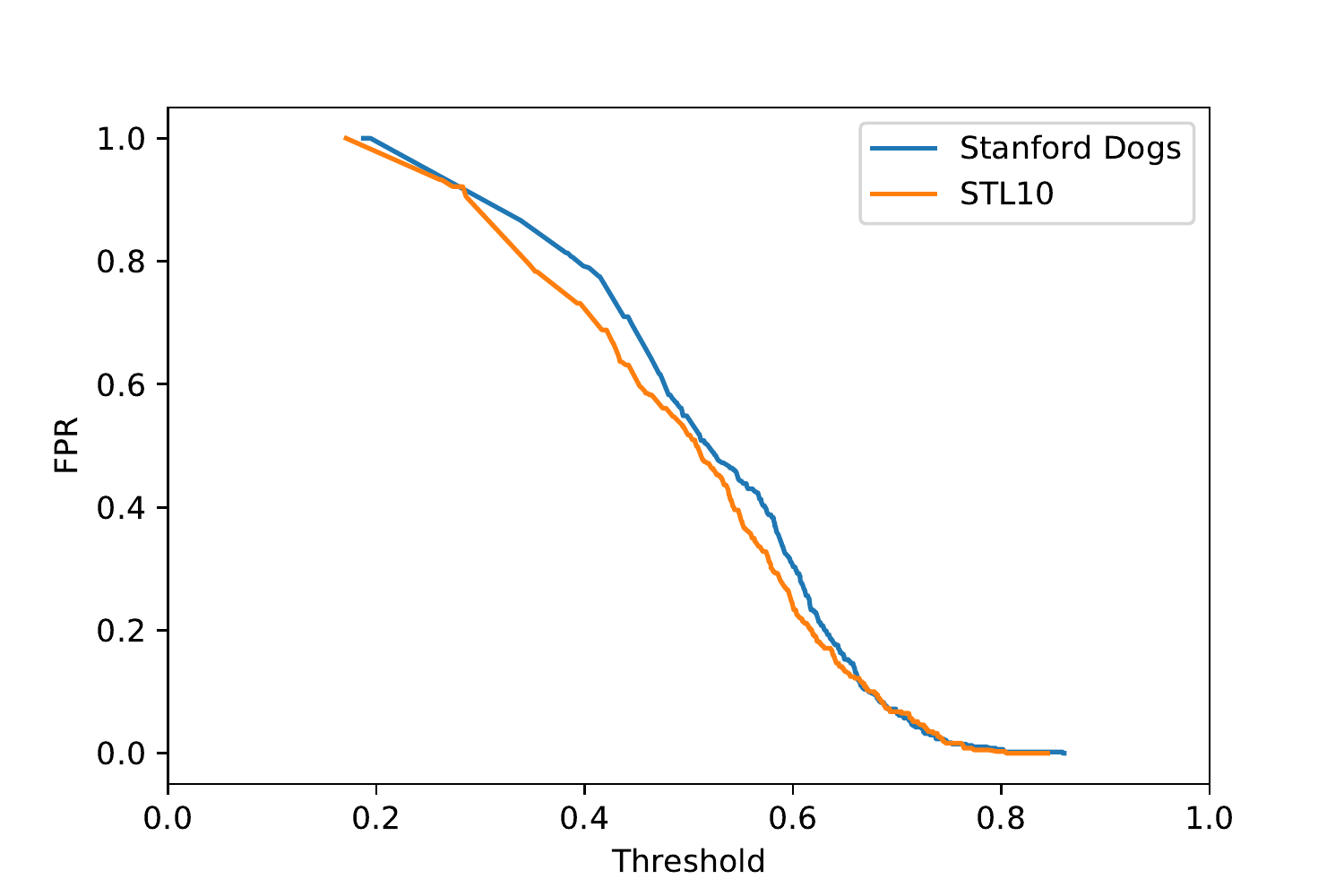} & \includegraphics[width=0.5\textwidth]{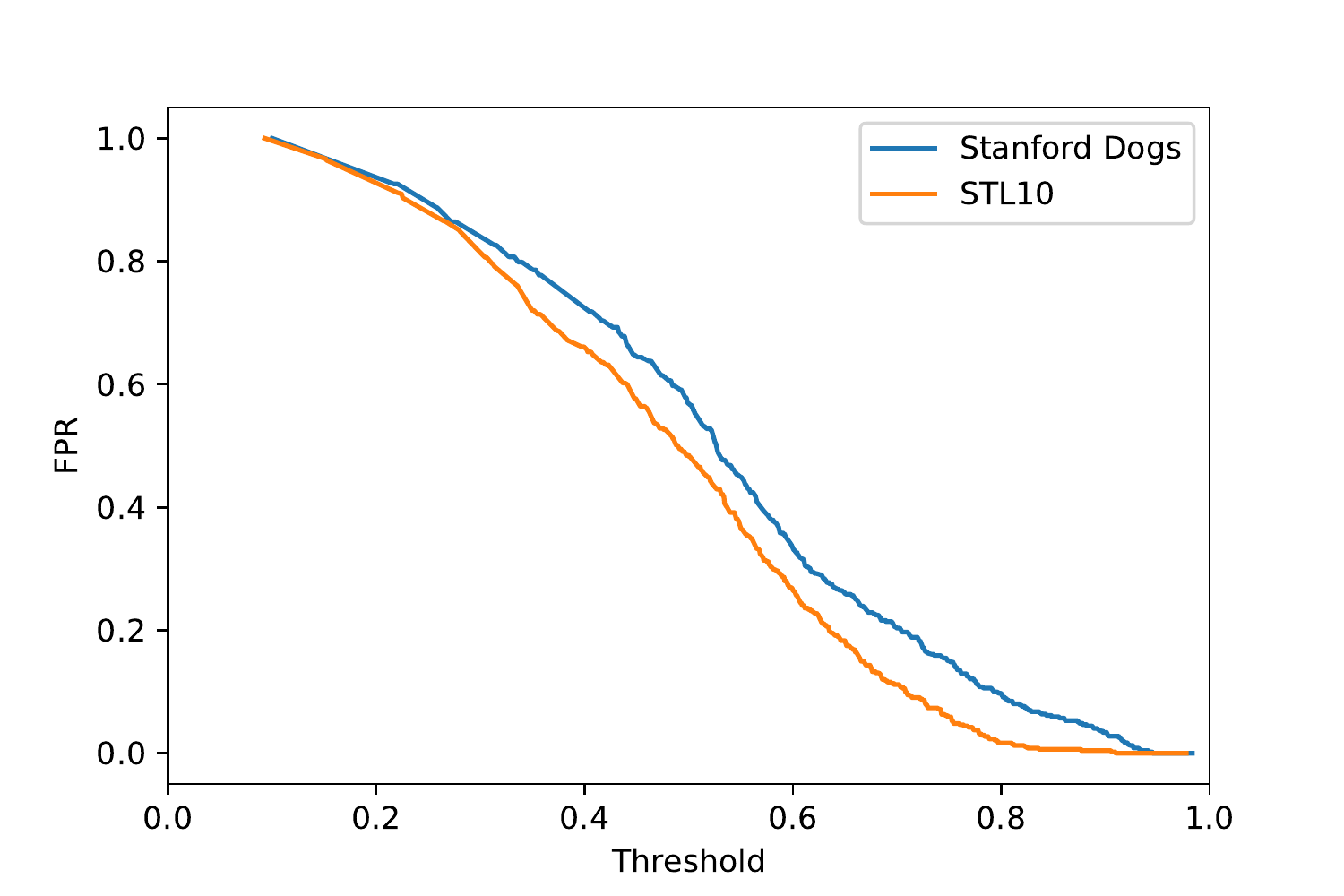}\\
    \end{tabular}
    \caption{False positive rate (FPR) as function of dSSIM (left) and dHaar (right) thresholds for the STL10 and Stanford Dogs datasets. We can observe that, in terms for FPR, the two metrics behave quite similarly, and most importantly, that they also behave very similarly across datasets.}
    \label{fig:fpr_correlation}
\end{figure*}

Figure~\ref{fig:ROC} shows the ROC curves of the Stanford Dogs dataset and STL-10 dataset, respectively. We observe that dSSIM and dHaar are the metrics that are the best correlated to human recognition ability. Both metrics achieve an AUC of at least 0.8, and up to 0.86 for both metrics depending on the dataset. In addition, Figure~\ref{fig:fpr_correlation} shows that the FPR of both the dSSIM and dHaar metrics are highly correlated across the two datasets, which indicates that the privacy threshold may generalize to other image datasets in a similar manner.

From the ROC curves, we can determine the best threshold that trades off FPR against TPR. We chose two popular methods~\cite{akobeng2007understanding} to select such a threshold: (i) by maximizing the accuracy (maximizing the difference TPR-FPR) and (ii) by choosing the cut-off threshold, i.e., the threshold mapped to the point on the ROC curve closest to the (0,1) corner. Table~\ref{table:thresholdsDogs} and Table~\ref{table:thresholdsSTL} show these thresholds,  $t_{\text{acc}}$ and $t_{\text{cut-off}}$ for the Stanford Dogs dataset and STL-10 dataset, respectively. We observe that the optimal thresholds are between 0.61 and 0.66 for dSSIM and 0.6 and 0.62 for dHaar. We also notice that, for a given metric, the higher the threshold, the smaller the FPR. This also implies that if we want to reduce the FPR and thus improve the privacy, we can also increase the obfuscation threshold. However, imposing a higher threshold would require stronger obfuscation parameters that might potentially harm the learning process.

\begin{table}[h!]
\centering
\begin{tabular}{|p{10mm} p{4mm} c c|} 
 \hline
 Metric & AUC & $t_{\text{acc}}$ (FPR, TPR) &  $t_{\text{cut-off}}$ (FPR, TPR)\\
 \hline
 MSE & 0.77 & 2178 (0.36, 0.79) & 2589 (0.31, 0.69) \\ 
 FID & 0.59 & 63160 (0.53, 0.69) & 93674 (0.43, 0.56)\\
 dSSIM & 0.80 & 0.63 (0.27, 0.77) & 0.65 (0.25, 0.74)\\
 pHash & 0.78 & 0.28 (0.25, 0.69) & 0.28 (0.25, 0.69)\\
 dHaar & 0.86 & 0.61 (0.23, 0.80) & 0.62 (0.21, 0.78)\\
 \hline
\end{tabular}
\caption{AUC and optimal thresholds for different metrics against humans on the Stanford Dogs dataset. $t_{\text{acc}}$ denotes the threshold that maximizes the accuracy and $t_{\text{cut-off}}$ the threshold picked by the cut-off method.}
\label{table:thresholdsDogs}
\end{table}

\begin{table}[h!]
\centering
\begin{tabular}{|p{10mm} p{4mm} c c|} 
 \hline
 Metric & AUC & $t_{\text{acc}}$ (FPR, TPR) &  $t_{\text{cut-off}}$ (FPR, TPR)\\
 \hline
 MSE & 0.70 & 2228 (0.45, 0.80) & 2965 (0.34, 0.65) \\ 
 FID & 0.65 & 415 (0.54, 0.75) & 34971 (0.40, 0.59)\\
 dSSIM & 0.86 & 0.66 (0.14, 0.72) & 0.61 (0.22, 0.77)\\
 pHash & 0.75 & 0.28 (0.27, 0.65) & 0.28 (0.27, 0.65)\\
 dHaar & 0.84 & 0.60 (0.23, 0.78) & 0.60 (0.22, 0.77) \\
 \hline
\end{tabular}
\caption{AUC and optimal thresholds for different metrics against humans on the STL-10 dataset. $t_{\text{acc}}$ denotes the threshold that maximizes the accuracy and $t_{\text{cut-off}}$ the threshold picked by the cut-off method.}
\label{table:thresholdsSTL}
\end{table}

\subsection{Computer Recognition}\label{sec:vision}

In this section, we evaluate if a computer can better recognize obfuscated images than humans, and ultimately understand the relation between the privacy metrics and the computer's ability to recognize the labels of obfuscated images. To do so, we rely on the state-of-the-art image classifier Google Vision AI.
The Vision API assigns labels to images along with a confidence score and classify them into about 20k predefined categories.

\begin{figure*}[t]
    \centering
    \begin{tabular}{c c}
    \includegraphics[width=0.5\textwidth]{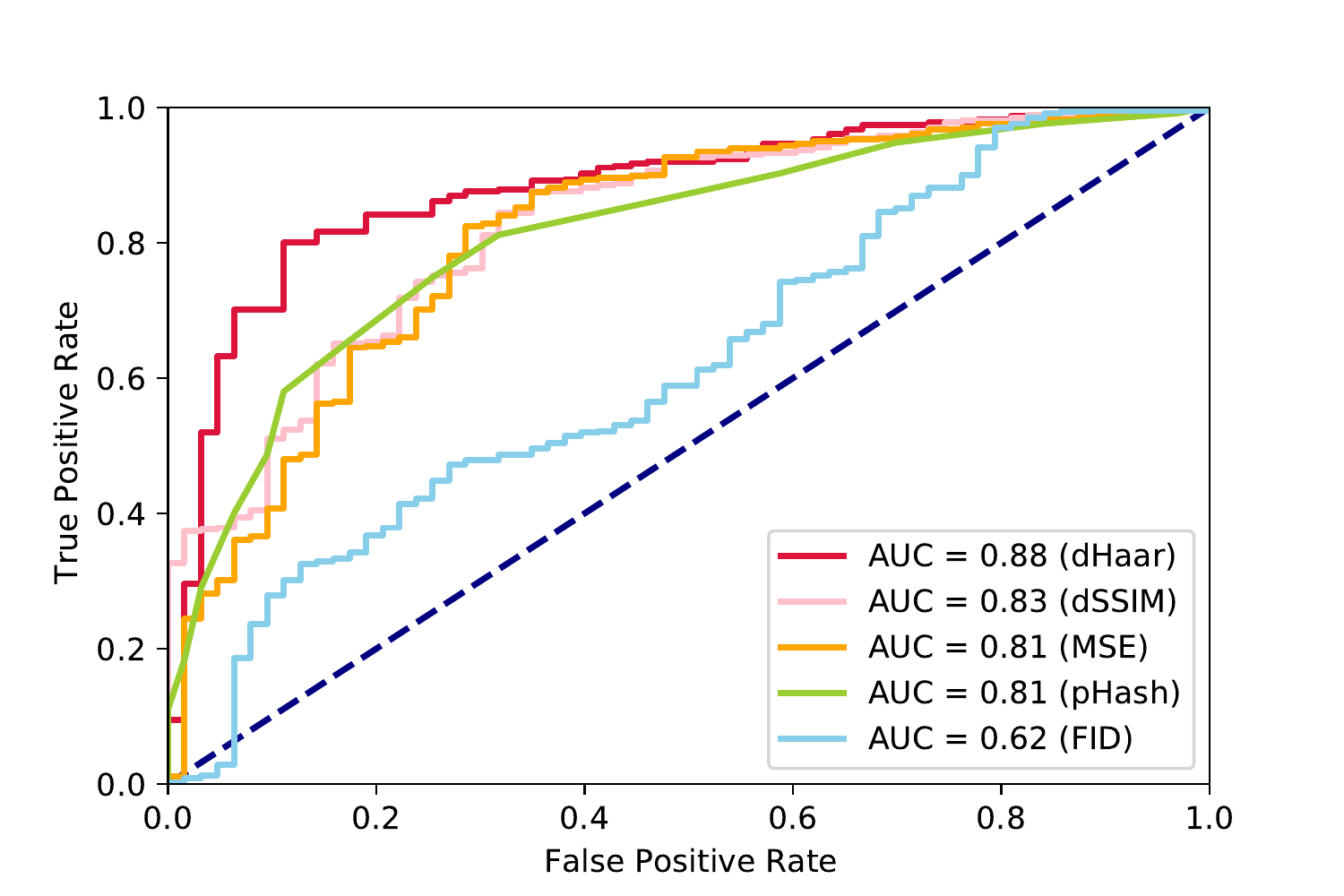} & \includegraphics[width=0.5\textwidth]{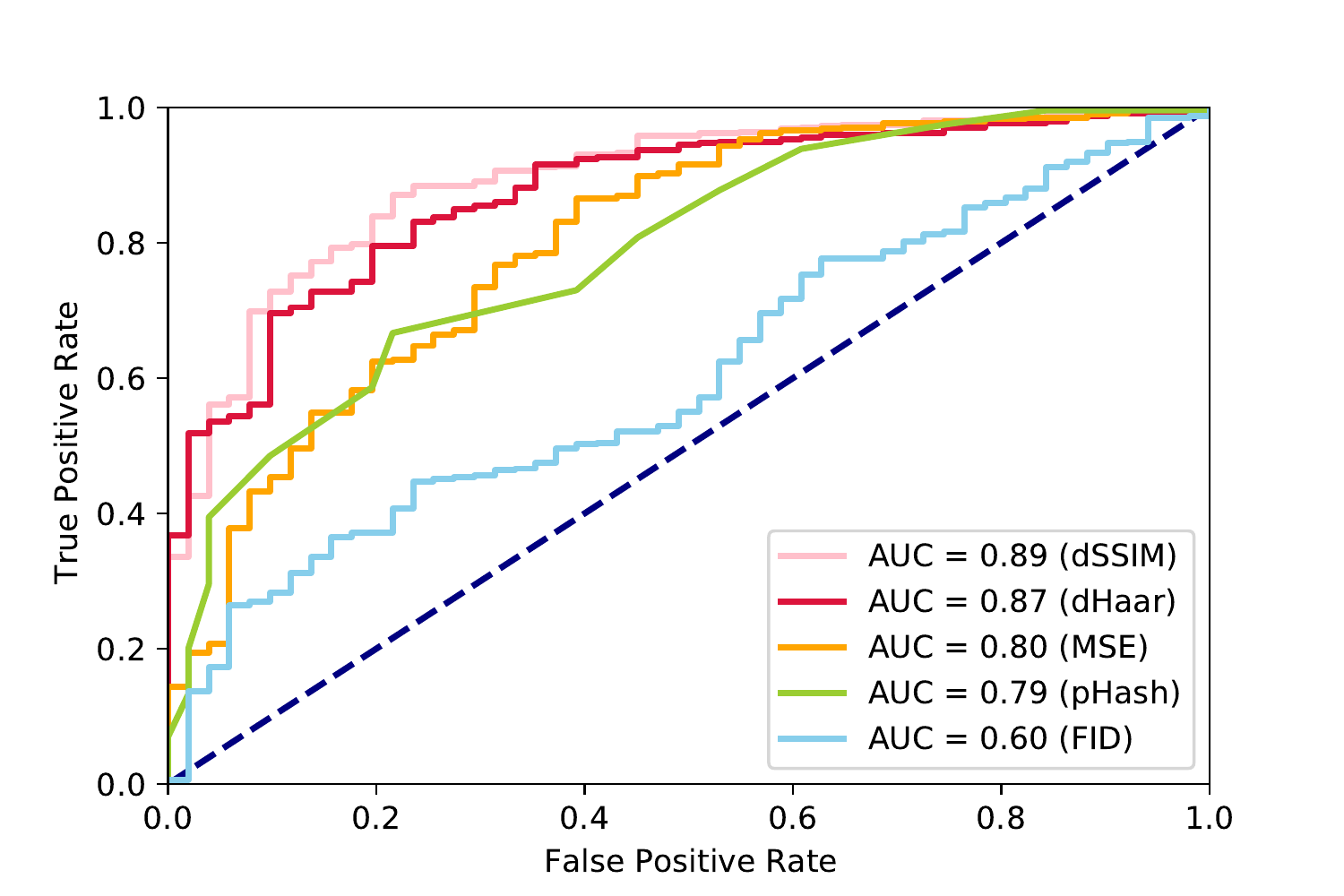}\\
    \end{tabular}
    \caption{ROC curves for different image quality metrics (IQM) against Google Vision on the Stanford Dogs dataset (left) and the STL-10 dataset (right). A higher curve indicates better agreement with human evaluations of the obfuscated images.  Again, as in Figure~\ref{fig:ROC}, the dHaar and dSSIM metrics show better agreement with human subjects as compared to other metrics.}
    \label{fig:ROC_google}
\end{figure*}

We use the same approach as for the human survey and query the Vision API with obfuscated images to evaluate if it can recognize their original labels. We construct the obfuscated images with the same parameters as in the user survey. In total, we send the equivalent of five surveys -- 245 images -- to the Vision API server for each dataset. Then we check if the classes of the original images are proposed in the top 10 labels inferred by Google Vision. If this is not the case, we conclude that the image is well obfuscated, else it is not. Following this, we plot the ROC curves shown in Figure~\ref{fig:ROC_google}. We first observe that dSSIM and dHaar are again the metrics that are best suited to predict Google Vision's performance.
We notice that the corresponding ROC curves have even better AUC values than for human recognition, meaning that these metrics are even better at predicting Google Vision's performance than human recognition performance. 

Looking at Tables \ref{table:thresholdsDogsGoogle} and \ref{table:thresholdsSTLGoogle}, we notice that the optimal thresholds for the Google Vision API (between 0.47 and 0.56 for dSSIM and between 0.54 and 0.55 for dHaar) are lower compared to those for human recognition. With these lower thresholds, the FPR is even smaller than in the case of human recognition. We further note that, by using the optimal human-level thresholds, we nearly halve the FPR with Google Vision (this is not in the tables). We conclude from this experiment that human-defined thresholds are stricter, and thus stronger from a privacy point of view. One explanation is that computers have more trouble at separating and recognizing mixed images than humans. This observation further indicates that the attack is hard to automatize and thus is not scalable for privacy parameters above human-level thresholds.

\begin{table}[h!]
\centering
\begin{adjustwidth}{-0.4cm}{}
\begin{tabular}{|p{10mm} p{4mm} c c|} 
 \hline
 Metric & AUC & $t_{\text{acc}}$ (fpr, tpr) &  $t_{\text{cut-off}}$ (fpr, tpr)\\
 \hline
 MSE & 0.81 & 1715 (0.28, 0.82) & 2118 (0.26, 0.72) \\ 
 FID & 0.62 & 100518 (0.26, 0.47) & 80275 (0.46, 0.53)\\
 dSSIM & 0.83 & 0.47 (0.34, 0.87) & 0.56 (0.25, 0.74)\\
 pHash & 0.81 & 0.21 (0.25, 0.74) & 0.21 (0.25, 0.74)\\
 dHaar & 0.88 & 0.56 (0.11, 0.80) & 0.55 (0.19, 0.81)\\
 \hline
\end{tabular}
\caption{AUC and optimal thresholds for tested metrics for Stanford Dogs and Google Vision. $t_{\text{acc}}$ denotes the threshold that maximizes the accuracy and $t_{\text{cut-off}}$ the threshold picked by the cut-off method.}
\label{table:thresholdsDogsGoogle}
\end{adjustwidth}
\end{table}

\begin{table}[h!]
\centering
\begin{adjustwidth}{-0.4cm}{}
\begin{tabular}{|p{10mm} p{4mm} c c|} 
 \hline
 Metric & AUC & $t_{\text{acc}}$ (fpr, tpr) &  $t_{\text{cut-off}}$ (fpr, tpr)\\
 \hline
 MSE & 0.80 & 1328 (0.39, 0.86) & 1850 (0.29, 0.73) \\ 
 FID & 0.60 & 181679 (0.23, 0.44) & 146856 (0.47, 0.52)\\
 dSSIM & 0.89 & 0.48 (0.21, 0.79) & 0.54 (0.19, 0.79)\\
 pHash & 0.79 & 0.25 (0.21, 0.66) & 0.25 (0.21, 0.66)\\
 dHaar & 0.87 & 0.54 (0.19, 0.79) & 0.54 (0.19, 0.79)\\
 \hline
\end{tabular}
\caption{AUC and optimal thresholds for tested metrics for STL10 and Google Vision. $t_{\text{acc}}$ denotes the threshold that maximizes the accuracy and $t_{\text{cut-off}}$ the threshold picked by the cut-off method.}
\label{table:thresholdsSTLGoogle}
\end{adjustwidth}
\end{table}

\subsection{Other Attack Vectors}\label{other_attacks}
We have so far studied two main methods to assess the robustness of our obfuscation schemes. However, there are other known de-obfuscation techniques that could be leveraged to reduce the effectiveness of these obfuscation schemes. In the following, we discuss them and clarify to what extent they can potentially jeopardize the proposed schemes.

\textbf{Unmixing.} 
Unmixing consists in recovering the original images $x_i$ and $x_j$ by knowing $\Tilde{x} = \lambda_i x_i + \lambda_j x_j$. This is similar to the problem of blind image separation (BIS). We want to make sure that our schemes are not prone to such BIS attacks. To do so, we do not reveal the mixing factors $\lambda_i$'s to the server and use each original image at most once per epoch. We hide the mixing factors by sharing only the label associated with the input image with strongest weight (as discussed in Section \ref{sec:obf}). 
The client performs the mixing locally, keeping $\lambda_i$'s hidden from the server. Unmixing would further first require to identify pairs of obfuscated images at different epochs sharing the same original image(s). Assuming there are $N$ obfuscated images per epoch, this would require $O(N!)$ comparisons. Last but not least, blind image separation is made even more complex by combining the mixing scheme with other obfuscation techniques, such as noising or shuffling, with different combinations and random values for different epochs.

\textbf{Interpolation.}
A risk specific to pixel grafting is to reconstruct one source image by interpolation. Indeed, if the two original images that are mixed and grafted do not have a similar distribution, there is a risk that the obfuscated image can be split into two distinct sparse images by identifying the pixels that have been grafted but not mixed (remember that one part of the image comes from the grafting operation and the rest from the mixing operation). Once the image is split into two sparse images, it becomes possible for the adversary to recover the image with grafted pixels by using interpolation techniques~\cite{S14}. The risk of separating the image into the two original sources can be reduced if the images come from similar distributions (e.g., if they are from the same class). However, we cannot assume universal security guarantees with the mixing and grafting technique. 

\textbf{Denoising.}
Denoising mechanisms could be used to reduce the effectiveness of the noise and mixing technique. To evaluate this threat,
we use OpenCV's Gaussian denoiser with recommended parameters\footnote{\url{https://docs.opencv.org/3.4/d1/d79/group__photo__denoise.html\#ga03aa4189fc3e31dafd638d90de335617}}. 
We apply the denoiser on 1000 randomly picked obfuscated images scoring over the privacy threshold 0.63, and compare the dSSIM score between the denoised images and the noisy images. We notice that the dSSIM score decreases by only 0.0001 for the denoised images, showing that of-the-shelf denoising techniques do not threaten our noise and mixing scheme.

\textbf{Deblurring.}
Similarly to denoising, deblurring mechanisms can be used to restore the quality of blurred content.
The Wiener deconvolution method~\cite{lim1990two} has widespread use in image applications.
It works in the frequency domain, attempting to estimate and reverse the convolution corresponding to the blurring operation. We apply Python Scipy's implementation of the Wiener deconvolution with default filter window size\footnote{\url{https://docs.scipy.org/doc/scipy/reference/generated/scipy.signal.wiener.html}} on images obfuscated with blurring and mixing.
The operation leads to a severe reduction of the dSSIM score. Even for images obfuscated with strong parameters, we observe that the privacy score is lowered by 43\% on average and drops to an average dSSIM score of 0.41, falling under our privacy thresholds.
A potential explanation is that the blurring and mixing operations are commutative, and thus applying blurring before mixing is equivalent to applying it after, which probably eases the deconvolution process.
We conclude that deblurring can significantly damage the privacy of our obfuscated images, and thus that blurring and mixing should not be considered more secure than simple mixing.

\textbf{CNN-based reconstruction attack~\cite{MSS16}.} 
The attack presented in~\cite{MSS16} aims to re-identify the labels of obfuscated targeted images by using a convolutional neural network (CNN) trained with labeled images obfuscated with the same technique as the targeted images. It is also assumed that the training images take labels in the same set as the targeted images. 
In order to assess the power of this attack against our obfuscation mechanisms, we train a CNN on an obfuscated STL10 dataset following the same architecture as~\cite{MSS16}.
The obfuscation parameters are chosen such that the privacy score of the dataset lies right above the threshold $t_\text{acc}$ found for the STL10 dataset in Section \ref{sec:user}. We report the CNN accuracy of our (robust) obfuscation methods in Table \ref{table:cnn}. We compare these results with pixelization with the same parameters as in the original study~\cite{MSS16}. Trained on data obfuscated with our obfuscation techniques, we show that it remains hard for the adversary to reconstruct the original images labels as in~\cite{MSS16}. Indeed, while the accuracy of pixelization (0.67) is  close to the (baseline) accuracy without any obfuscation (0.73), the risk with our obfuscation mechanisms is dramatically reduced, down to 0.19 with shuffling and mixing, which is much closer to random guessing (0.1) than any previously proposed image obfuscation technique.

\begin{table}
  \centering
  \begin{tabular}{|l | l |} \hline
    Technique & Accuracy \\ \hline
    None & 0.73\\
    Pixelization & 0.67\\
    Pix. and mixing & 0.23\\
    Shuffling and mixing & 0.19\\ 
    Noise and mixing & 0.21 \\ \hline
  \end{tabular}
  \caption{Accuracy of the deep learning-based reconstruction attack from~\cite{MSS16} against the STL10 dataset obfuscated with various techniques. None represents the accuracy on the non-obfuscated data, and pixelization is depicted as a point of comparison (as it was already shown to be insecure in~\cite{MSS16}).}
  \label{table:cnn}
\end{table}

\subsection{Discussion}\label{sec:bss}

Privacy metrics are in general hard to find, especially regarding human visual perception. Despite that, we find two existing image quality metrics that match human perception and return similar values among two different datasets. Besides, we define a procedure to determine the privacy threshold of any dataset based on user surveys. It is to be noted that the metrics we consider are meant to measure degradation of quality and not the degree of privacy as such. However, we re-purpose these metrics for quantifying privacy since the privacy we introduce in images via obfuscation is based on the degradation of the content of the image.

Our adversarial assumptions are relatively strong since our thresholds are upper bounds on safe values to release images. This is for two main reasons. First, we picked images that are relatively easy to recognize, and whose labels are clearly distinct from one another. However, this would be certainly more challenging with image sets where domain expertise is required, e.g., in medical images such as MRIs, and more generally when fine-grained details need to be discovered that would allow re-identification of some an individual or some sensitive attribute. Furthermore, in our survey, we provide the possible target labels beforehand, which dramatically reduces the complexity of the recognition task. In general, no prior would be given to the adversary. So, generally speaking, we may assume that other non-natural images categories, such as medical images, can be privacy-protected with much less obfuscation, using parameters aiming at a much lower threshold. A user survey of domain experts may have to be carried out to empirically discover the best thresholds according to our proposed methodology.

\section{Exploring the Privacy-Utility Trade-off}\label{sec:results}

We have shown how to quantify privacy introduced via obfuscation. We now show how to use this ability to explore the privacy versus utility trade-off.

\subsection{Experimental Setup}

In order to assess the utility of our obfuscation techniques, we train a model on a dataset obfuscated with each technique, and measure its accuracy on the same, non-obfuscated, test set. We compare the resulting accuracy values with the reference accuracy given by a model trained on non-obfuscated data.
Indeed, for the purpose of fitting real-life applications such as MLaaS, we expect the model to accurately classify non-obfuscated images, representing realistic clients' prediction queries. 

We train our models on a binary image classification task.
The Cats vs Dogs~\cite{catdogs} dataset provides a large enough collection of natural images that fit our task.
The labeled dataset is split into 22,800 training images and 2000 test images, all resized to 256 by 256 pixels. We use a residual neural network, specifically the ResNet18 model proposed in~\cite{HZR+16}.
The model is trained for 20 epochs, uses batches of 200 images containing equal number of dot and cat images, and Adam~\cite{adam} optimizer with default parameters. It is to be noted that in our experiments, the goal is not to obtain the maximum possible accuracy for a given dataset but to compare the performance between obfuscated and non-obfuscated data under exactly similar conditions.

We rely on an iterative training process where, at each epoch, the client freshly obfuscates its whole dataset and shares it with the server.
The server can update its model and wait for the next mixture (obfsucated dataset) to start a new epoch. The resulting model can either be kept in the cloud or sent back to the client. Mixtures are randomly generated for each epoch with the obfuscation techniques defined in Section 3, and parameters chosen according to the privacy level desired by the client. Note that we keep the same technique for all the epochs of a given model, but that images are randomly assigned to other images at each epoch, and similarly for the other mechanisms (e.g., a new noise is drawn at random for every image and every epoch). We try two mixing approaches. In the first one, the client creates the mixture by blindly mixing its dataset with a permutation of it elements without looking at the labels. 
In a second approach, we explore a constrained mixing where the client mixes samples coming from the same class only, i.e., dog with dogs and cats with cats in our case. We refer to the latter as intra-class mixing.

We tested the performance of all obfuscation methods presented in Section~\ref{sec:obf} but report here only the robust ones. Thus, we exclude the results of the combined application of mixing and grafting. We include two previously proposed mechanisms, i.e., blur and pixelization, for comparison. We also compare the combined effect of blurring and mixing with that of blurring only. As shown in Table~\ref{table:acc}, we test our obfuscation mechanisms with different parameters and both classical and intra-class mixing, to provide suggestions for practical use. We report the results for the most interesting parameters for each obfuscation technique.
The model -- thus baseline accuracy -- without any obfuscation technique is referred to as Reference in the table.
The privacy score is measured using the dSSIM metric.

\begin{table*}
\begin{tabular}{ l | l | l | l | l | l  }
  \hline
 \textbf{Technique} & \textbf{Parameters} &\textbf{Privacy} &\textbf{Privacy (intra)} & \textbf{Accuracy (in \%)} & \textbf{Accuracy (intra, in \%)} \\   \hline\hline
 Reference & - & 0 & 0 & 93.35 & 93.35 \\
   \hline
Simple mixing & $\lambda = 0.75$ & 0.14 & 0.15 & 92.75 & 92.45\\
 Simple mixing & $\lambda = 0.6$ & 0.29 & 0.28 & 87.65 & 90.50\\
 Simple mixing & $\lambda = 0.5$ & 0.32 & 0.30 & 79.45 & 79.20\\ \hline
  Noise and mixing & $\lambda = 0.75, \sigma$ = 20 & 0.65 & 0.68 & 91.30 & 88.94\\
    Noise and mixing & $\lambda = 0.5, \sigma$ = 20 & 0.72 & 0.71 & 69.00 & 81.85\\
 Noise and mixing & $\lambda = 0.75, \sigma$ = 30 & 0.80 & 0.79 & 85.55 & 86.60\\
  Noise and mixing & $\lambda = 0.5, \sigma$ = 30 & 0.81 & 0.83 & 71.20 & 79.20\\ \hline
    Shuffling and mixing & $\lambda = 0.75, b = 4$ & 0.46 & 0.47 & 88.35 & 84.85\\
  Shuffling and mixing & $\lambda = 0.5, b = 4$ & 0.56 & 0.56 & 68.40 & 85.30\\
  Shuffling and mixing & $\lambda = 0.75, b = 8$ & 0.63 &  0.63 & 80.45 & 84.95\\
    Shuffling and mixing & $\lambda = 0.5, b = 8$ & 0.70 & 0.70 & 69.20 & 83.90\\
  \hline
  Pix. and mixing & $\lambda = 0.75, s = 32$ & 0.52 & 0.52 & 80.10 & 79.90\\
Pix. and mixing  & $\lambda = 0.5, s = 32$ & 0.56 & 0.57 & 76.05 & 78.90\\
Pix. and mixing  & $\lambda = 0.5, s = 16$ & 0.57 & 0.61 & 65.20 & 67.50\\
Pix. and mixing & $\lambda = 0.75, s = 16$ & 0.59 & 0.59 & 67.70 & 67.40\\
 
    \hline
\end{tabular}
\caption{Accuracy and privacy on the cats vs. dogs classification task with various obfuscation techniques and their most relevant parameters.}
\label{table:acc}
\end{table*}

\subsection{Results}

\begin{figure}[t]
    \centering
    \includegraphics[width=1.07\columnwidth]{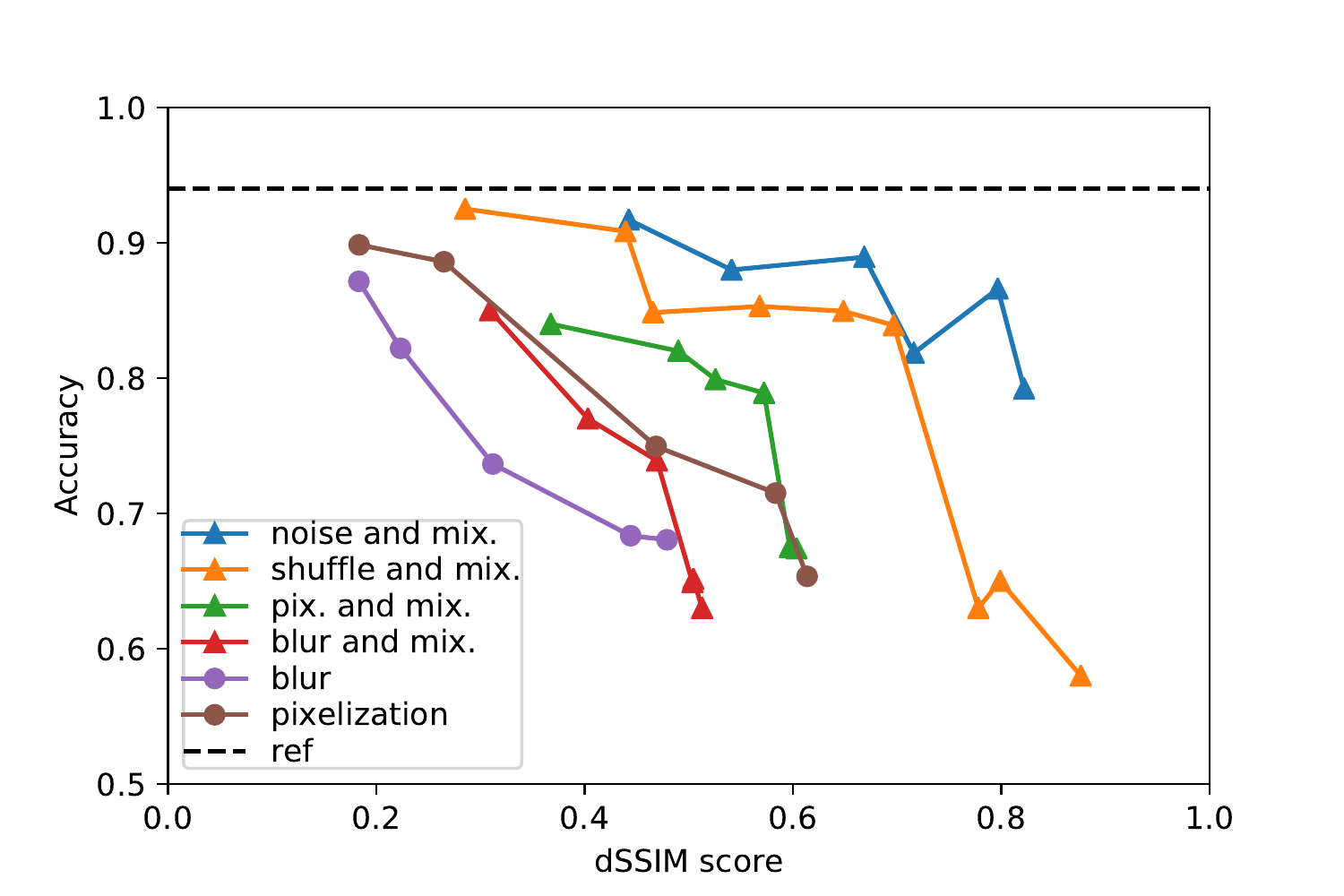}
    \caption{Privacy vs. accuracy for the proposed obfuscation techniques (constrained on intra-class mixing). The dashed black line represents the reference accuracy, i.e., the one calculated on the original, non-obfuscated, data. Blur and pixelization are included as comparison points.}
    \label{fig:pirvacc}
\end{figure}

Table \ref{table:acc} first shows that, in most cases, for similar parameters and privacy, intra-class mixing provides a higher test accuracy for similar privacy levels.
In addition to Table \ref{table:acc}, we plot the results of various tests of our obfuscation chemes against naive approaches (namely blurring and pixelization) for intra-class mixing (Figure~\ref{fig:pirvacc}).
Ideally, we want the obfuscation technique to maximize both privacy and utility. As we can see in Figure~\ref{fig:pirvacc}, the curve closest to the ideal curve is the combination of adding noise and mixing.
It provides, for good privacy values (larger than the thresholds defined in Table~\ref{table:thresholdsDogs}), an accuracy only 2 percentage points, respectively 4 percentage points, smaller than the reference accuracy for unconstrained and intra-class mixing, respectively. As comparison, the most related work by Lee et al. shows a decrease in accuracy of about 20 percentage points compared to the reference accuracy~\cite{lee2019synthesizing}.

As an additional remark, we notice that mixing images helps augment the accuracy even when combined with other forms of obfuscation.
When mixing is combined with blurring or pixelization, the resulting model is more accurate on test data that when performing obfuscation with blurring or pixelization alone. In general, we notice that the three obfuscation schemes that are robust to various attack vectors (including the one of McPherson et al. \cite{MSS16}) discussed in Section~\ref{other_attacks} are also those that provide the best privacy-utility trade-off.

Please note that all our obfuscation operations are linear in the number of pixels and thus have a very low complexity (in the order of a few milliseconds per image of size $256\times 256$). Additionally, training a single epoch of our networks takes about 43 seconds on the GPU and this time is obviously agnostic to the type of obfuscation that is applied to the input image. It takes the same amount of time as training on the original images.
\section{Related Work}\label{sec:related}

In this section, we cover three categories of the related literature: (i) image obfuscation schemes, (ii) attacks against obfuscation schemes, and (iii) generic methods for training machine-learning models in a privacy-preserving manner.

\textbf{Image Obfuscation.}
Lee et al. first proposed to build upon mixup and release random linear combinations of the original images for privacy-preserving machine learning. They combine up to 128 samples together and add noise in a differentially private manner~\cite{lee2019synthesizing,LLK+18}. However, they do not derive any metric to quantify the actual obfuscation brought by the random mixing, thus relying only on the differential privacy to ensure the privacy of the released dataset. Therefore, the accuracy of their ML model on their differentially private data is at best $\approx 20\%$ worse than the baseline accuracy. 
Huang et al. also build upon mixup to increase privacy in distributed machine learning~\cite{huanginstahide}. They show that mixing images alone is insecure, especially when public images are mixed with their private dataset, and they propose to randomly flip pixel signs to make it more private.  However, the flipped pixels are easily detectable due to color saturation, and potentially recoverable through interpolation~\cite{S14}, thus jeopardizing their security guarantees. Besides, their setting is different than ours as they consider distributed clients that only share their gradients and not the obfuscated images themselves.

Other schemes, not based on mixup, were previously proposed for image obfuscation.
P3~\cite{RGO13} was introduced as a privacy-preserving photo sharing system that encrypts the significant coefficients in the JPEG representation of the image. P3 aims to make the image unrecognizable yet preserve its JPEG structure and enable servers to perform certain operations on it. Ilia et al. rely on blurring to prevent Facebook users to from identifying other users in images with restricted access~\cite{IPA+15}.
In the same context of online social networks, He et al. propose to encrypt specific sensitive regions of an image (e.g., faces) for privacy-preserving image sharing~\cite{HLK+16}.
Sun et al. present a head inpainting method that is more robust to automatic person recognizers than common techniques such as blurring and that generates more natural patterns~\cite{SMO18}.
Poller et al. propose to rely on permutation in the spatial domain (i.e., shuffling) and masking as image obfuscation methods~\cite{PSL12}. None of these other schemes aim to preserve the utility of the data in a machine-learning context.

Fan uses pixelization (referred to as mosaicing) and Laplace noise as obfuscation. She extends the differential privacy concept to image data by defining a new neighborhood notion based on the difference in the pixels of two images~\cite{F18, F19}. She shows that her approach is able to prevent the reconstruction attacks via deep learning~\cite{MSS16} but at the cost of image quality degradation. Moreover, we believe that the differential privacy-inspired metric is not adapted to capture privacy in images as a simple shift might mislead the metric into high privacy values. 

Regarding privacy metrics, Newton et al. build on the concept of $k$-anonymity to de-identify faces in video surveillance images~\cite{NSM03}.
Tekli propose a framework for evaluating the robustness of image obfuscation techniques against deep learning-based reconstruction attacks~\cite{TBC+19}. To do so, they use two image metrics, among which SSIM like us. However, they use it as a similarity measure to evaluate the quality of the reconstruction attack of McPherson et al.~\cite{MSS16} against the proposed obfuscation mechanisms (pixelization, blurring and masking). They do not use it to quantify image privacy under any attack scenario, including a visual one. Fu et al. rely on pHash to evaluate the quality of adversary reconstruction~\cite{FWX+19}. But again, it is only used to evaluate one reconstruction attack and not to quantify image privacy in a more general case.

\textbf{Breaking Image Obfuscation.}
As discussed in Section~\ref{other_attacks}, McPherson et al. show that it is possible to recover information from images hidden by P3, blurring, or pixelization by using deep learning~\cite{MSS16}.
Wilber et al. show that Facebook face detector can break various image obfuscation techniques, such as blurring or ``privacy glasses''~\cite{WSB16}.
Solutions obfuscating only regions of interest, such as faces, can also be recovered based on background or contextual information.
For instance, Oh et al. rely on other pictures shared on social media to recognize individual faces even if they are fully obfuscated in the targeted pictures~\cite{OBF+16}.
Neustaedter et al. demonstrate how blurring fails to preserve privacy for home-based video conferencing~\cite{NGB06}.

Hill et al. recover sequences of characters from text obfuscated with pixelization and blurring~\cite{HZS+16}.
Lander et al. test the effectiveness of blurring and pixelization with user experiments~\cite{LBH01}. They ask participants to identify famous people in obfuscated movie clips and static images and show that they can still be recognized. Gopalan et al. show that subspace representation can help recognize faces obfuscated with blurring~\cite{GTT+12}. Punnappurath et al. extend this work to handle non-uniform motion blur, and illumination and pose variations~\cite{PRT+15}.
Gross et al. \cite{GSC+06} re-identify faces from blurred and pixelated face images and propose a new approach based on $k$-anonymity~\cite{NSM03} that preserves better privacy and data utility.

\textbf{Privacy-Preserving ML Training.}
Generic approaches have been proposed for training ML models in a privacy-preserving manner which can be categorized in two groups: (i) cryptography-based methods, and (ii) differential privacy (DP)-based methods.
Most cryptography-based approaches rely on homomorphic encryption and are limited to linear models or logistic regression models~\cite{aono2016scalable,bonte2018privacy, crawford2018doing,giacomelli2018privacy,graepel2012ml,jiang2018securelr,kim2018logistic,kim2018secure}. Only a very few consider more complex models such as deep neural networks \cite{aono2017privacy,nandakumar2019towards}. However, such advanced models have to compromise on the accuracy of the model or on the training cost which remains prohibitive given current computational capabilities. Moreover, cryptographic approaches require a substantial change of the ML backend architecture.

Regarding DP-based methods, Abadi et al. consider a centralized setting where a \emph{trusted} party holds the data, trains the ML model, and performs the noise addition~\cite{ACG+16}.
On the contrary, other works propose to rely on distributed learning~\cite{CDN+15,huang2019dp,SS15}. They consist in training local ML models on local data and then share only the parameters relevant to the global model with the central server (e.g., the gradients), typically by adding noise in a differentially private manner.
However, training accurate ML models generally requires high privacy budgets, thus questioning the privacy achievable in practice~\cite{jayaraman2019evaluating}.

\section{Conclusion}\label{sec:conclusion}
We introduce a method for quantifying visual privacy introduced in images through obfuscation. We are able to do this by assessing existing full-reference image quality metrics for their agreement with human perception through a user study. We further validate our privacy metrics with a state-of-the-art computer vision system. These metrics then help directly in choosing the best privacy versus utility trade-off for a given machine-learning application. To the best of our knowledge, we are the first to address jointly address the challenges of quantifying image privacy and of optimizing the privacy-utility trade-off in deep learning-based systems where the adversary has access to the individual training samples.

We address the case of image classification by deep learning models, when the labels of the images are not privacy-sensitive. This leaves room to confirm that our methodology can be extended to the use of other signals such as speech, music, or text, as well as other machine learning problems like regression, image segmentation, speech (and other signal) generation etc. Our work may also inspire the development of novel quality metrics that are dedicated to assess human perception in images, speech, music, and text.  


\appendix

\section{Appendix: Survey}
\label{appendix:survey}

We provide here more information about our surveys. The questionnaire filled by the participant is introduced with the following paragraph:

\textit{In this survey you are asked to recognize the content of images.
The survey contains 49 questions and should take maximum 10 minutes to complete. The difficulty varies through the questions, so it is normal that some are harder than others.
You can zoom in and out, but cannot apply any further modifications to the images (such as contrast modification) in order to help you guess the labels. The images may contain several objects, so you may check several boxes, but do this only if you think that all objects are individually present. Do not tick several boxes if you hesitate between those.
If you don't know, use the ``I cannot tell`` option.}

For the Stanford Dogs-based survey, an annex helping users with the breeds was added along with:

\textit{You can find on the ref.pdf document the list of breeds that you can select from, along with an example. No other breed is going to appear.}

\begin{figure*}[h]
    \centering
    \begin{tabular}{c c}
    \includegraphics[width=0.48\textwidth]{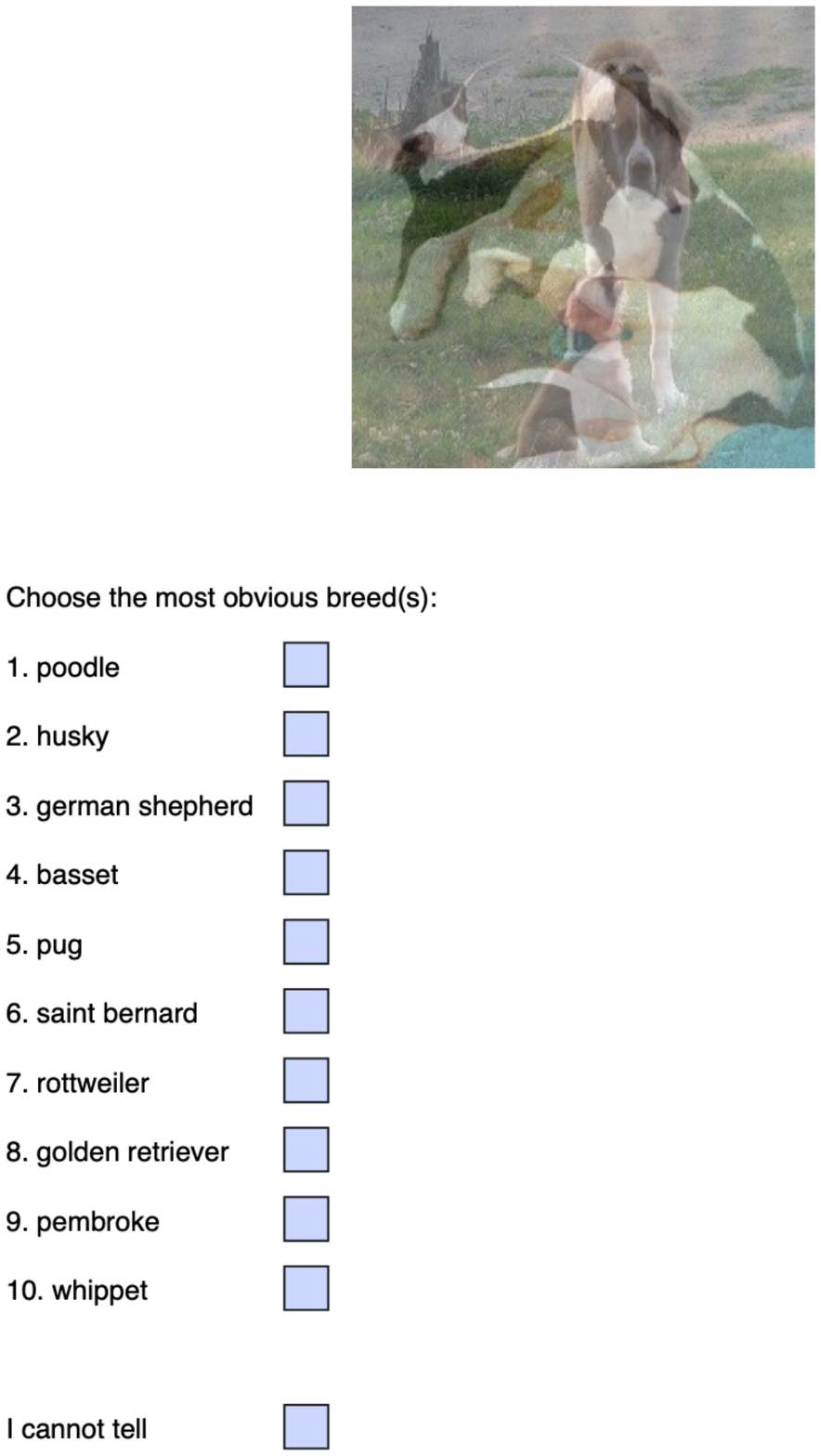} & \includegraphics[width=0.48\textwidth]{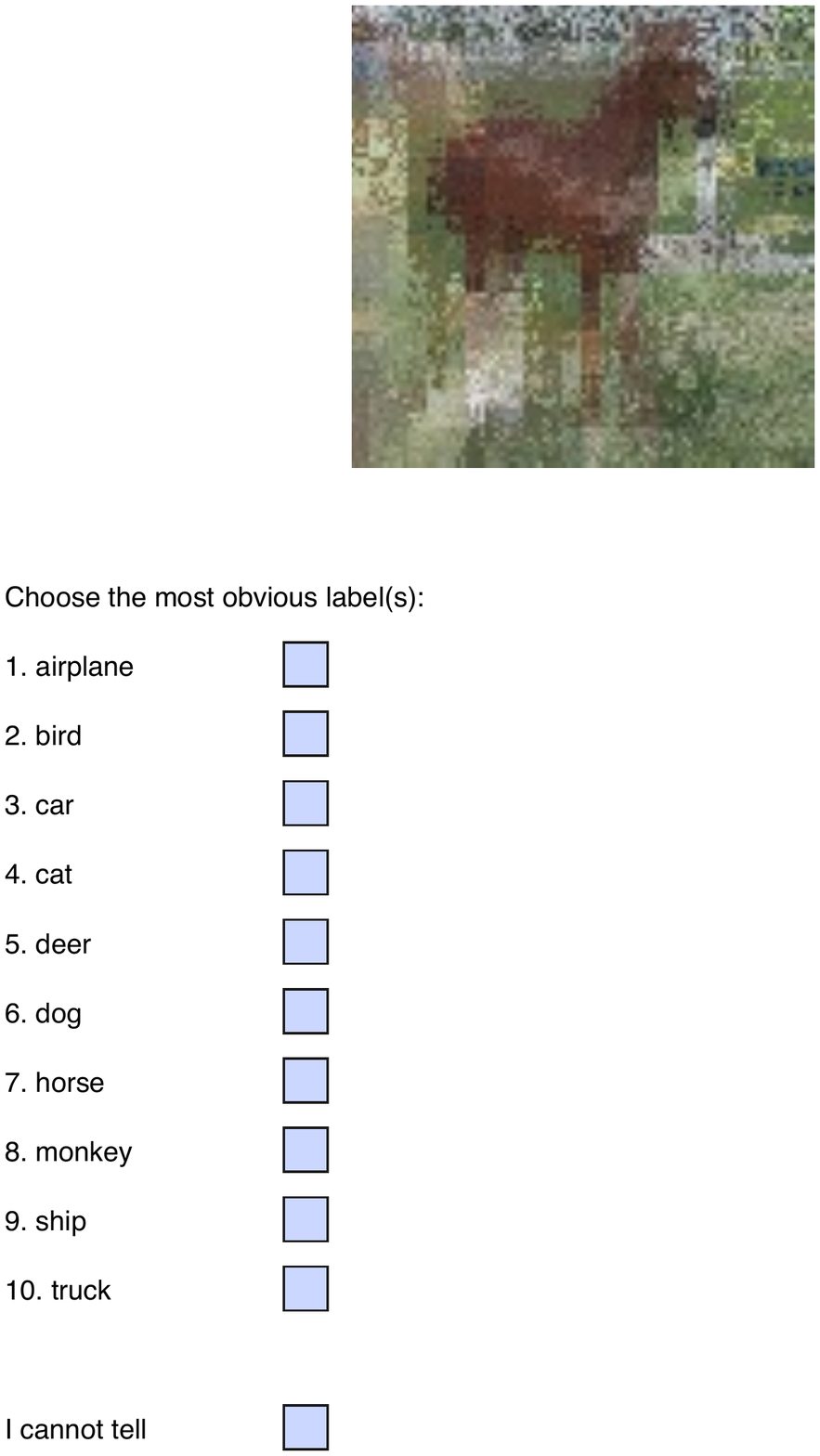}\\
    \includegraphics[width=0.48\textwidth]{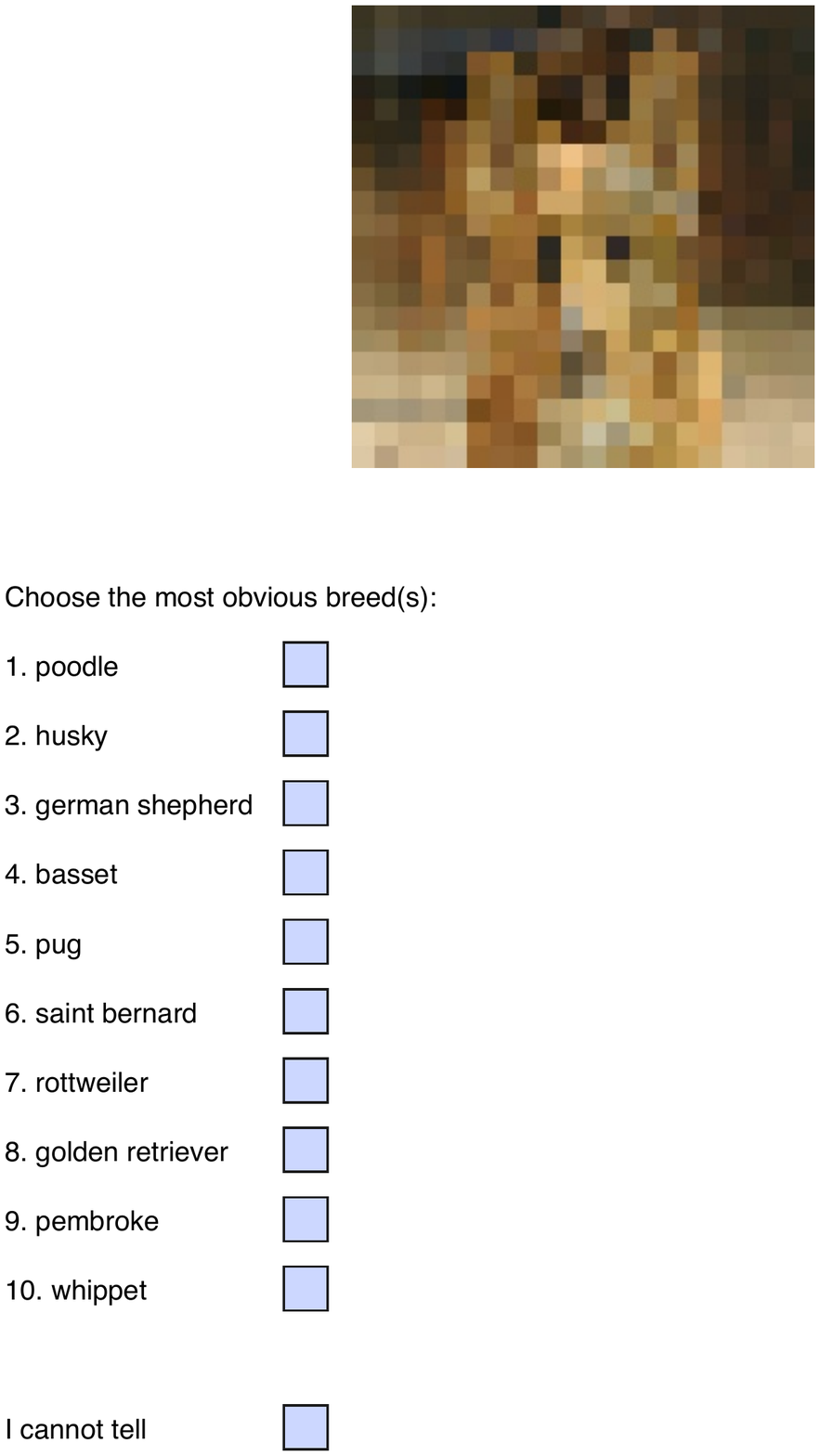} & \includegraphics[width=0.48\textwidth]{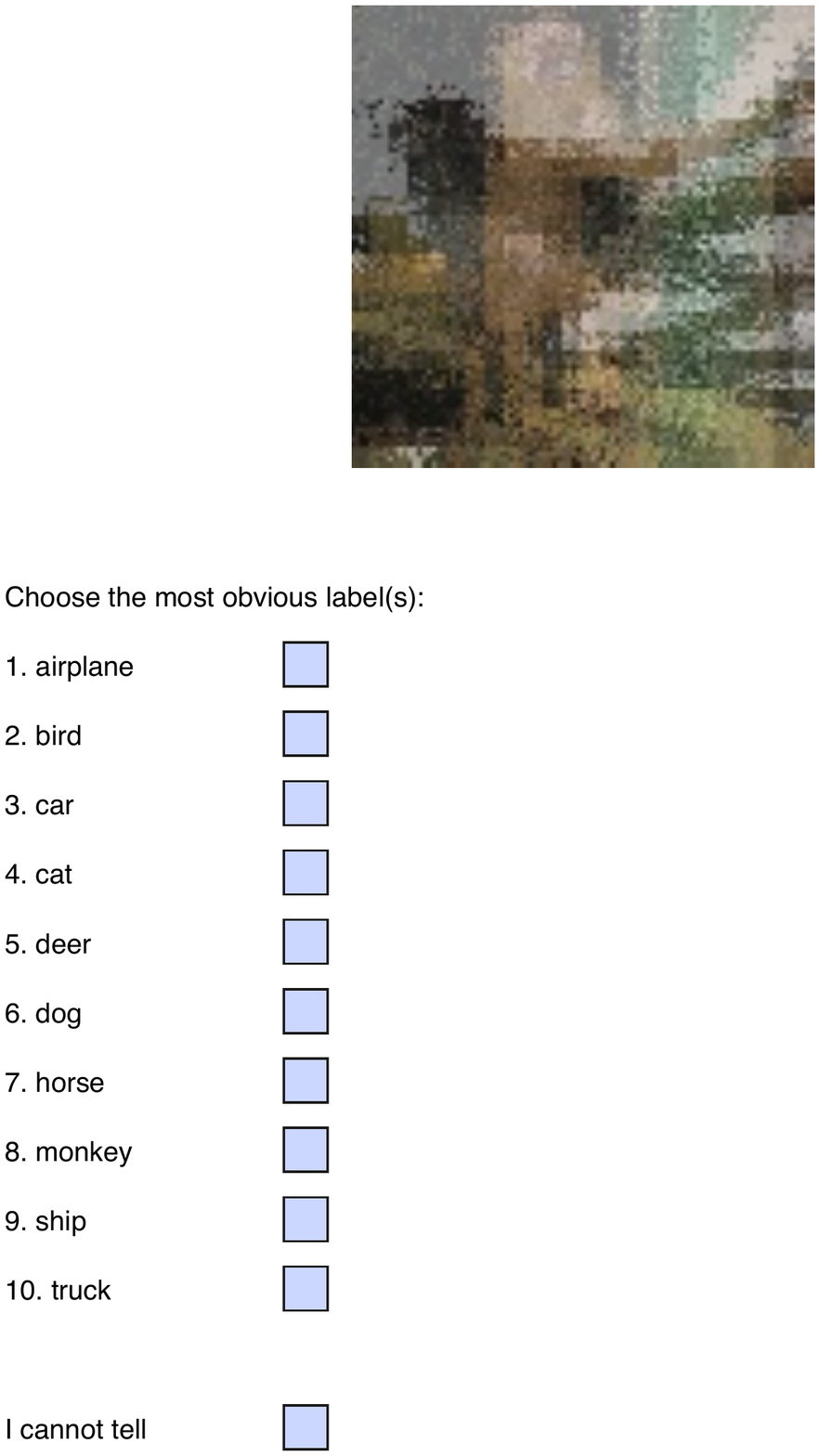}\\
    \end{tabular}
    \caption{Sample pages taken from surveys generated from the Stanford Dogs dataset (left) and the STL10 dataset (right).}
    \label{fig:exsPage}
\end{figure*}

Sample pages extracted from surveys given to the participants can be found in Figure~\ref{fig:exsPage}.

\end{document}